\documentclass[twocolumn,prd,showpacs]{revtex4}
\usepackage{graphicx}

\begin{document}
\renewcommand{\theequation}{\thesection.\arabic{equation}}
\newcommand{\re}{\mathop{\mathrm{Re}}}
\newcommand{\be}{\begin{equation}}
\newcommand{\ee}{\end{equation}}
\newcommand{\bea}{\begin{eqnarray}}
\newcommand{\eea}{\end{eqnarray}}
\title{Observational Constraints on finite scale factor singularities}
\author{Tomasz Denkiewicz}
\email{atomekd@wmf.univ.szczecin.pl}
\affiliation{\it Institute of Physics, University of Szczecin, Wielkopolska 15,
          70-451 Szczecin, Poland}
\affiliation{\it Copernicus Center for Interdisciplinary Studies,
S{\l }awkowska 17, 31-016 Krak\'ow, Poland}
\date{\today}
\input epsf
\begin{abstract}
We discuss the combined constraints on a Finite Scale Factor Singularity (FSF) universe evolution scenario, which come from the shift parameter $\mathcal{R}$, baryon acoustic oscillations (BAO) $\mathcal{A}$, and from the type Ia supernovae.
We show that observations allow existence of such singularities in the $2\times10^9$ years in future (at $1\sigma$ CL) which is much farther than a Sudden Future Singularity (SFS), and that at the present moment of the cosmic evolution, one cannot differentiate between cosmological scenario which allow finite scale factor singularities and the standard {\bf $\Lambda$CDM} dark energy models. We also show that there is an allowed value of $m=2/3$ within $1\sigma$ CL, which corresponds to a dust-filled Einstein-de-Sitter universe limit of the early time evolution and so it is pasted ito a standard early-time scenario.
\end{abstract}


\maketitle

\section{Introduction}
\setcounter{equation}{0}
Finite scale factor singularities (FSF) are one of types of new and exotic singularities which were found first in Ref. \cite{BGT}. The inspiration to search these new types of singularities was due to the observations of high-redshift type Ia supernovae (SNIa) which provided strong evidence that the expansion of the universe is accelerating due to an unknown form of dark energy\cite{supernovaeold} phenomenologically behaving as the cosmological constant.
Further observational data \cite{supernovaenew} made cosmologists think of more accelerating universe filled with phantom \cite{phantom} which violated all energy conditions: the null ($\varrho c^2 + p
\geq 0$), weak ($\varrho c^2 \geq 0$ and $\varrho c^2 + p \geq 0$),
strong ($\varrho c^2 + p \geq 0$ and $\varrho c^2 + 3p \geq 0$), and dominant energy ($\varrho c^2 \geq 0$, $-\varrho c^2 \leq p \leq
\varrho c^2$) ($c$ is the speed of light, $\varrho$ is the mass density in $kg m^{-3}$ and $p$ is the pressure). Phantom-driven dark energy leads to a big-rip singularity (BR or type I according to \cite{nojiri}) in which the infinite values of the energy density and pressure ($\rho$, $p\to\infty$) are accompanied by the infinite value of the scale factor ($a\to\infty$) \cite{caldwellPRL}.

The list of new types of singularities contains: a big-rip (BR), a sudden future singularity (SFS) \cite{barrow04}, which can appear in inhomogeneous and anisotropic models too \cite{barrow042, sfs1}, a generalized sudden future singularity (GSFS), a finite scale factor singularity (FSF) \cite{aps}, a big-separation singularity (BS). They are characterized by violation of all or some of the energy conditions which results in a blow-up of all or some of the physical quantities: the scale factor, the energy density and the pressure. The finite scale factor singularity, which is the subject of this paper, is a weak singularity according to Tipler's definition, but strong according to Kr\'{o}lak's definition \cite{lazkoz}. Apart from mentioned above there are also $w$-singularities \cite{wsin}, which are not physical singularities, but are singularities of a barotropic index $w$ which are present in different cosmological models of $f(R)$ gravity \cite{star1980}, in scalar field models \cite{Setare}, and in brane cosmologies \cite{Sahni}.

For a finite scale factor singularity (FSF) $\rho \to \infty$ and $p \to \infty $ diverge, while the scale factor $a$ remains constant. This means that it is similar to a big-bang singularity with only one exception - the scale factor is a constant $a = a_s=$ const instead of zero $a_{BB} = 0$. Besides, FSF singularities are stronger than a sudden future singularities (SFS) [10], so they placed themselves in between the big-bang (which is strong and geodesically incomplete) and the SFS. It has been shown that SFS appear in physical theories such as $f(R)$ gravity, scalar field cosmologies, brane cosmologies \cite{yuri} and, in particular, they plague loop quantum cosmology \cite{LQC}. It seems then that after an appropriate choice of the scalar field, $f(R)$ function or brane, FSF will be easily deduced to appear in such physical theories as well.

\indent The paper is organized as follows. In section \ref{s2} we present FSF scenario. In section \ref{oc} we derive the expressions for the observables: baryon acoustic oscillations, distance to the last scattering surface, type Ia supernovae, used to test an FSF scenario. In section \ref{rac} we give the results and discussion.

\section{Finite Scale Factor Singularity scenario construction}\label{s2}

In order to obtain an FSF singularity one should consider the simple
framework of an Einstein-Friedmann cosmology governed by the
standard field equations
\bea \label{rho} \varrho(t) &=& \frac{3}{8\pi G}
\left(\frac{\dot{a}^2}{a^2} + \frac{kc^2}{a^2}
\right)~,\\
\label{p} p(t) &=& - \frac{c^2}{8\pi G} \left(2 \frac{\ddot{a}}{a} + \frac{\dot{a}^2}{a^2} + \frac{kc^2}{a^2} \right)~,
\eea
where the energy-momentum conservation law
\be
\label{conser}
\dot{\varrho}(t) = - 3 \frac{\dot{a}}{a}
\left(\varrho(t) + \frac{p(t)}{c^2} \right)~,
\ee
is trivially fulfilled due to the Bianchi identity. Here $a \equiv a(t)$ is the scale factor, the dot means the derivative with respect to time $t$, $G$ is the gravitational constant, and the curvature index $k=0, \pm 1$. During further considerations we set $c=1$ and $k=0$.

Similarly like in the case of SFS, which were tested against the observations in \cite{DHD, GHDD, DDGH}, one is able to obtain an FSF singularity by taking the scale factor in the form
\be \label{sf2} a(y) = a_s \left[\delta + \left(1 - \delta \right) y^m -
\delta \left( 1 - y \right)^n \right]~, \hspace{0.5cm} y \equiv \frac{t}{t_s} \ee
with the appropriate choice of the constants $\delta, t_s, a_s, m,
n$. In contrast to an SFS in order to have accelerated expansion of the universe $\delta$ has to be positive  ($\delta>0$). For $1<n<2$ we have an SFS,  which plagues the loop quantum cosmology \cite{LQC}. In order to have an FSF singularity instead of SFS, $n$ has to lie in the range $0<n<1$.

As can be seen from (\ref{rho})-(\ref{sf2}), for an FSF $\rho$ diverges and we have: for $t\rightarrow t_s$, $a\rightarrow a_s$, $\rho\rightarrow\infty$, and $|p|\rightarrow\infty$, where $a_s,\ t_s,\ \rho_s$, are constants and $a_s\neq 0$.

In a model, expressed in terms of the
scale factor (\ref{sf2}), the evolution begins with the standard big-bang
singularity at $t=0$ for $a=0$, and finishes at a finite scale factor singularity for $t=t_s$
where $a=a_s\equiv a(t_s)$ is a constant. In terms of the rescaled time $y$ we have $a(1) = a_s$.

The standard Friedmann limit (i.e. models without a singularity) of
(\ref{sf2}) is achieved when $\delta \to 0$; hence $\delta$ is called
the ``non-standardicity" parameter. Additionally,
notwithstanding Ref. \cite{barrow04} and in agreement with the field
equations (\ref{rho})-({\ref{p}), $\delta$ can be
both positive and negative leading to an acceleration or a
deceleration of the universe,
respectively.\\
\indent It is important to our discussion that the asymptotic behaviour of the scale factor (\ref{sf2}) close to the big-bang singularity at $t=0$ is given by a simple power-law $a_{\rm BB} = y^m$, simulating the behaviour of flat $k=0$ barotropic fluid models with $m = 2/[3(w+1)]$~.\\
\indent  The FSF singularity scenario consists of two dominating components such as a nonrelativistic matter, and the fluid which is driving the singularity. We consider the case of the noninteracting components from which both of them obey independently their continuity equations as \ref{conser}.
The evolution of both ingredients is independent. Nonrelativistic matter scales as $a^{-3}$
\be
\rho_m=\Omega_m\rho_0\left(\frac{a_0}{a}\right)^3
\ee
and the evolution of the other fluid, which we name here $\rho_Q$, can be determined by taking the difference between whole energy density, $\rho$, evolution from Friedmann eq. \ref{rho} and the
 $\rho_m$:
\be
\rho_Q=\rho-\rho_m
\ee
this ingredient, $\rho_Q$, of the content of the Universe is responsible for the singularity in the energy density for $t\rightarrow t_s$.

\subsection{Classical stability of FSFS with respect to small perturbations}
\indent In Ref. \cite{JBSZWL} the study of classical stability of sudden singularities and ``Big Rip''  singularities with respect to small inhomogeneous scalar, vector and tensor perturbations was given. The analysis was made in the framework of gauge-invariant formalism introduced by Mukhanov \cite{MUK2}. Characterization of sudden singularities in terms of the series expansion of the scale factor on the approach to a singularity was given, and it was shown that once the density diverges near to a singularity it is unstable under small scalar perturbations. On the other hand, sudden singularities are stable to inhomogeneous vector and (tensor) gravitational-wave perturbations. Taking the Friedmann equations (\ref{rho})-(\ref{p}) with $c=1$, $8\pi G=1$ and the scale factor in terms of the series expansion on the approach to a singularity:
\begin{equation}
  a (t) = c_0 + c_1 (t_s - t)^{\lambda_{}} + \ldots, \label{apser}
\end{equation}

\begin{equation}
\dot{a} (t) = - c_1 \lambda_{} (t_s - t)^{\lambda - 1} + \ldots, 
\end{equation}

\begin{equation}
\ddot{a}(t)=c_{1}\lambda _{{}}(\lambda -1)(t_{s}-t)^{\lambda _{{}}-2}+\ldots,
\end{equation}%
were the $c_i$, $\lambda$ are real constants, with $c_i\neq0$, $c_0>0$, in the limit $t\rightarrow t_s$ the approximation for the density and pressure was derived \cite{JBSZWL}:

\begin{equation}
\rho = \frac{3 K}{c_0^2} + \frac{3 \lambda^2 c_1^2}{c_0^2} (t_s - t)^{2
(\lambda - 1)} + \ldots, 
\end{equation}

\begin{equation}
p = - \frac{K}{c_0^2} - \frac{2 c_1 \lambda (\lambda - 1)}{c_0} (t_s -
t)^{\lambda - 2} + \ldots \label{ppser}
\end{equation}

The density diverges if and only if $\lambda<1$, and the pressure diverges, if $\lambda<2$. Substituting (\ref{apser}-\ref{ppser}) into the gauge-invariant scalar perturbations equation \cite{MUK2}: 
\bea
a^2 \ddot{\Phi} + (4 + 3 p^{\prime}(\rho)) a \dot{a}^{} \dot{\Phi} +\nonumber \\ (2 a 
\ddot{a} + ( \dot{a}^2 - k) (1 + 3 p^{\prime}(\rho)) + p^{\prime}(\rho) K^2)
\Phi = 0,  \label{scalarpert}
\eea
for plane wave perturbations with wave number $K$, where $p'(\rho)\equiv dp/d\rho$, and further neglecting higher-order terms, and taking the case of $\lambda<1$ we have \cite{JBSZWL}:

\begin{equation}
\ddot{\Phi} - (\lambda - 2) T^{- 1} \dot{\Phi} + \frac{c_1}{c_0} \lambda^2
T^{\lambda - 2} \Phi = 0,
\end{equation}
where $T=t_s-t$ and dots indicate differentiation with respect to $t$. By transformation of the variables, one can express this equation as the Bessel equation, and then obtain its solution to be \cite{JBSZWL}: 

\begin{equation}
\Phi (t) = A (t_s - t)^{\lambda - 1} + B,
\end{equation}
where $A$, $B$ are constants. For FSFS with $\lambda<1$, $\Phi(t)$ diverges as $t\rightarrow t_s$. Such a result is valid for all wavelengths, and for all values of $k$.\\ 
 
\section{Observational constraints}\label{oc}
\setcounter{equation}{0}

We consider three observational constraints, such as luminosity distance moduli to type Ia supernovae, baryon acoustic oscillations, and the shift parameter which is scaled distance to the last scattering surface of the cosmic microwave background.
We make use of Markov Chain Monte Carlo technique within the framework of Bayesian statistics to obtain posterior probability distribution as a function of the parameters: $n$, $m$, $\delta$, $y_0=\frac{t_0}{t_s}$, where $t_0$ is the present age of the universe. We employ Metropolis-Hastings algorithm with uniform priors: $\delta\in(0,30)$, $n\in(0,1)$, $m\in(0,3)$, $y_0\in(0,1)$.
The results are marginalised over $H_0 = 74.2 \pm 3.6$ taken from HST Key Project.

\subsection{Supernovae}
We proceed within the framework of Friedmann cosmology, and consider an observer located at $r=0$ at coordinate time $t=t_0$. The
observer receives a light ray emitted at $r=r_1$ at coordinate time
$t=t_1$. We then have a standard null geodesic equation
\be
\label{geod}
\int_0^{r_1} \frac{dr}{\sqrt{1-kr^2}} = \int_{t_1}^{t_0}
\frac{cdt}{a(t)} = c t_s \int_{y_1}^{y_0}
\frac{dy}{a(y)}~,
\ee
with the scale factor $a(y)$ given by (\ref{sf2}). Using (\ref{sf2}) again, the redshift is given by
\be
\label{redshift}
1+z=\frac{a(t_0)}{a(t_1)} = \frac{\delta +
\left(1 - \delta \right) y_0^m - \delta \left( 1 - y_0 \right)^n}
{\delta + \left(1 - \delta \right) y_1^m - \delta \left( 1 - y_1
\right)^n}~,
\ee
where $y_0 = y(t_0)$ and $y_1 = y(t_1)$.

The luminosity distance to supernova is given by
\be
d_L(z)=(1+z)c t_s \int_{y_1}^{y_0}
\frac{dy}{a(y)}
\ee
and the distance modulus is
\be
\mu(z)=5\log_{10}d_L(z)+25.
\ee
The $\chi^2$ for the SNIa data is
\be
\chi^2_{SN}=\sum^{N}_{i=1}\frac{(\mu_{obs}(z_i)-\mu(z_i))^2}{\sigma^2_i+\sigma_{int}^2},
\ee
where $\sigma_i$ is the quoted observational error on the $i^{\rm th}$ Union2 SNIa and $\sigma_{\rm int}$ is the SNIa intrinsic scatter. Following
\cite{Amanullah} we take $\sigma_{\rm int}=0.15$.
\subsection{Shift parameter}
\label{shift}
The standard formula for the CMB shift parameter is given by
\cite{bond97,Nesseris:2006er}:
\be
\label{shift}
{\cal R}=\frac{l_1^{\prime TT}}{l_1^{TT}}~,
\ee
where $l_1^{TT}$ is the temperature perturbation CMB spectrum multipole of the first acoustic peak in the model under consideration and $l_1^{\prime TT}$ corresponds to a reference flat standard Cold Dark Matter (CDM) model. The multipole number is related to an angular scale of the sound horizon $r_s$ at decoupling by \cite{PRD41,singh03}
\be
\label{theta}
\theta_1 = \frac{r_s}{d_A} \propto \frac{1}{l_1}~,
\ee
where
\be
r_s = a_{\rm dec} S(r_s) = a_{\rm dec} S\left( \int_{0}^{t_{\rm dec}} c_s \frac{dt}{a(t)} \right)~,
\ee
with $c_s$ being the sound velocity and the angular diameter distance reads as
\be
d_A = a_{\rm dec} S(r_{\rm dec}) = a_{\rm dec} S\left( \int_{t_{\rm dec}}^{t_0} c \frac{dt}{a(t)} \right)
\ee
where $c$ is the velocity of light and $S(r) = r$ for $k=0$.

Following \cite{Nesseris:2006er} we can write, using (\ref{shift}) and (\ref{theta}) as follows
\be
{\cal R} = \frac{r_s}{r'_s} \frac{d'_A(z'_{\rm dec})}{d_A(z_{\rm dec})}~,
\ee
which, by assuming that at decoupling the amount of radiation was the same in both the flat reference standard CDM model and in our FSF model (which we assume to be just the same as a standard matter-radiation model of the early universe, since FSF models do not change the evolution there) we have that
\be
\frac{r_s}{r'_s} = \frac{1}{\sqrt{\Omega_{m0}}}~.
\ee
On the other hand, for a reference standard CDM model
\bea
d'_A &=& \frac{2c a'_{\rm dec}}{a_0H_0} \left[\sqrt{\Omega'_{r0} + 1}
- \sqrt{\Omega'_{r0} + \frac{a'_{\rm dec}}{a_0}} \right] \nonumber \\
&=& \frac{2c a'_{\rm dec}}{a_0H_0} f(\Omega'_{r0},a'_{\rm dec})~,
\eea
while for our FSF model the angular diameter distance is given by
\be
d_A = a_{\rm dec} r_{\rm dec}
\ee
with $r_{\rm dec}$ given by
taken at decoupling.
Using the above, we may write that for our FSF models the shift parameter is
\be
{\cal R} =
\frac{2c}{H_0a_0 \sqrt{\Omega_{m0}} r_{\rm dec}} = \frac{2ct_s}{a'(y_0) \sqrt{\Omega_{m0}} r_{\rm dec}}~,
\ee
where we have assumed that the function $f(\Omega'_{r0},a'_{\rm dec})$ is approximately unity \cite{Nesseris:2006er}.

Finally, the rescaled shift parameter is
\bea
\bar{\mathcal{R}}&=&\sqrt{\Omega_{m0}} \frac{H_0a_0}{c} r_{\rm dec} = \sqrt{\Omega_{m0}} \int_0^z\frac{dz^\prime}{E(z^\prime)} \nonumber \\
&=& \sqrt{\Omega_{m0}} a'(y_0) \int_{y_{\rm dec}}^{y_0} \frac{dy}{a(y)},
\eea
where $E(z)=H(z)/H_0$.
The WMAP data gives $\bar{\mathcal{R}}=1.725\pm0.018$ \cite{7y},  thus $\chi^2$ for the shift parameter is the following
\be
\chi^2_R=\frac{(\bar{\mathcal{R}}-1.725)^2}{0.018^2}~.
\ee
\subsection{Baryon acoustic oscillations}

In Ref. \cite{eisenstein} it was suggested that instead of taking the position of an acoustic peak one should measure the large-scale correlation function at $100 h^{-1}$ Mpc separation using the sample of 46748 luminous red galaxies (LRG) selected from the SDSS (Sloan Digital Sky Survey) sample. The appropriate quantity to be measured is known as the ${\cal A}$ parameter and reads as
\be
{\cal A}= \Omega_{0m}^{1/2}E(z_{BAO})^{-1/3}\left[\frac{1}{z_{BAO}}\int^{z_1}_0\frac{dz}{E(z)} \right]^{2/3}~.
\ee

According to Ref. \cite{eisenstein} ${\cal A}$ should have the value
\be
{\cal A} = 0.469 \left( \frac{n}{0.98} \right)^{-0.35} \pm 0.017~~,
\ee
where $n$ is the spectral index (now taken about $\sim 0.96$).\\
We have $\chi^2$ for BAO
\be
\chi^2_A=\frac{({\cal A}-0.469)^2}{0.017^2}~.
\ee
Using (\ref{sf2}) and (\ref{redshift}) we get the form of ${\cal A}$ for an FSF as follows
\bea
{\cal A}=a^{\prime}(y_0)\left[\frac{a(y_{bao})}{a^{\prime}(y_{bao})a(y_0)}\right]^{\frac{1}{3}}\left[\frac{1}{z_{bao}}\int_{y_{bao}}^{y_0}\frac{dy}{a(y)}\right]~.
\eea
\begin{figure*}
    \begin{tabular}{ccc}
      \resizebox{50mm}{!}{\rotatebox{-90}{\includegraphics{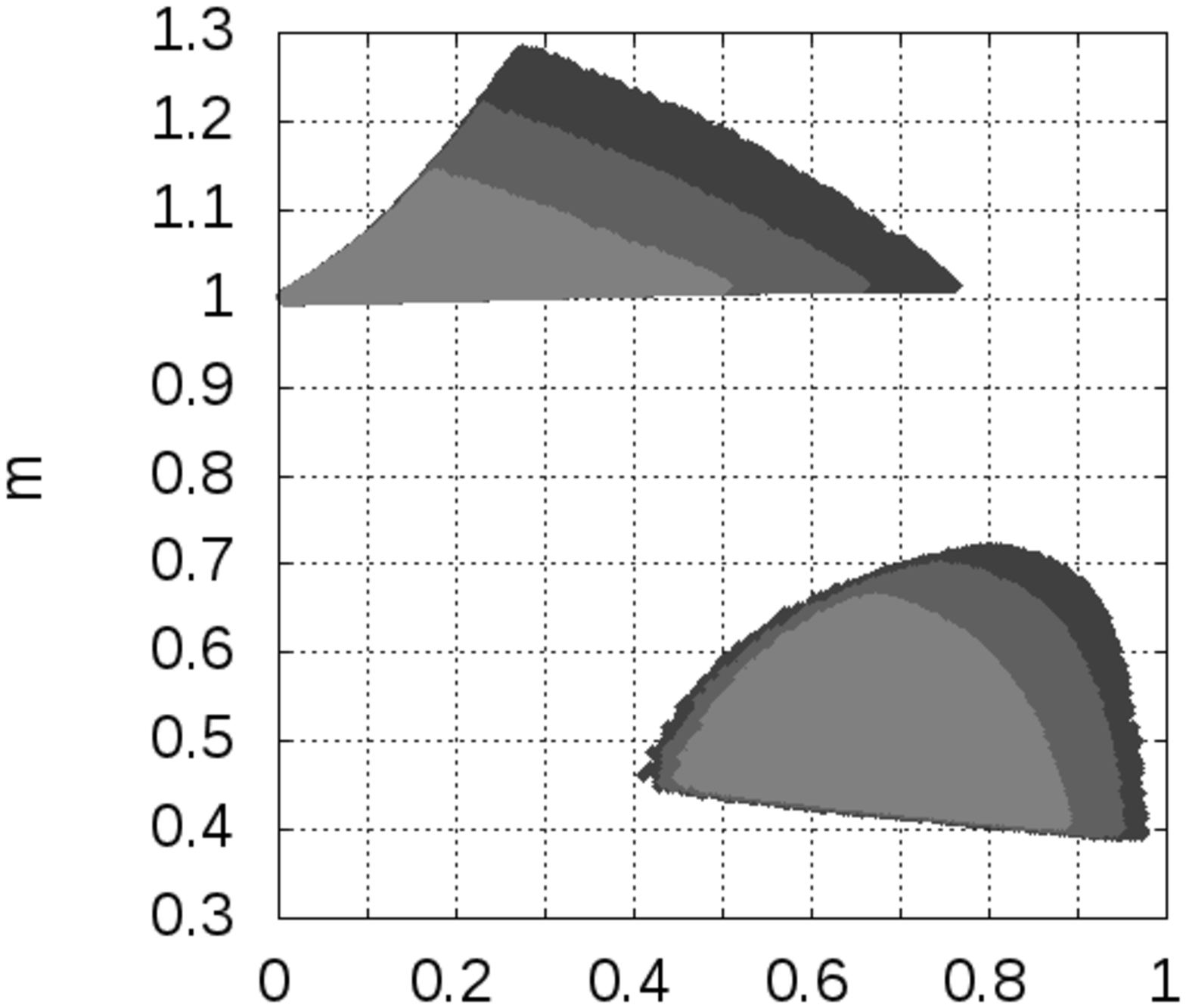}}} &
       &
       \\

      \resizebox{50mm}{!}{\rotatebox{-90}{\includegraphics{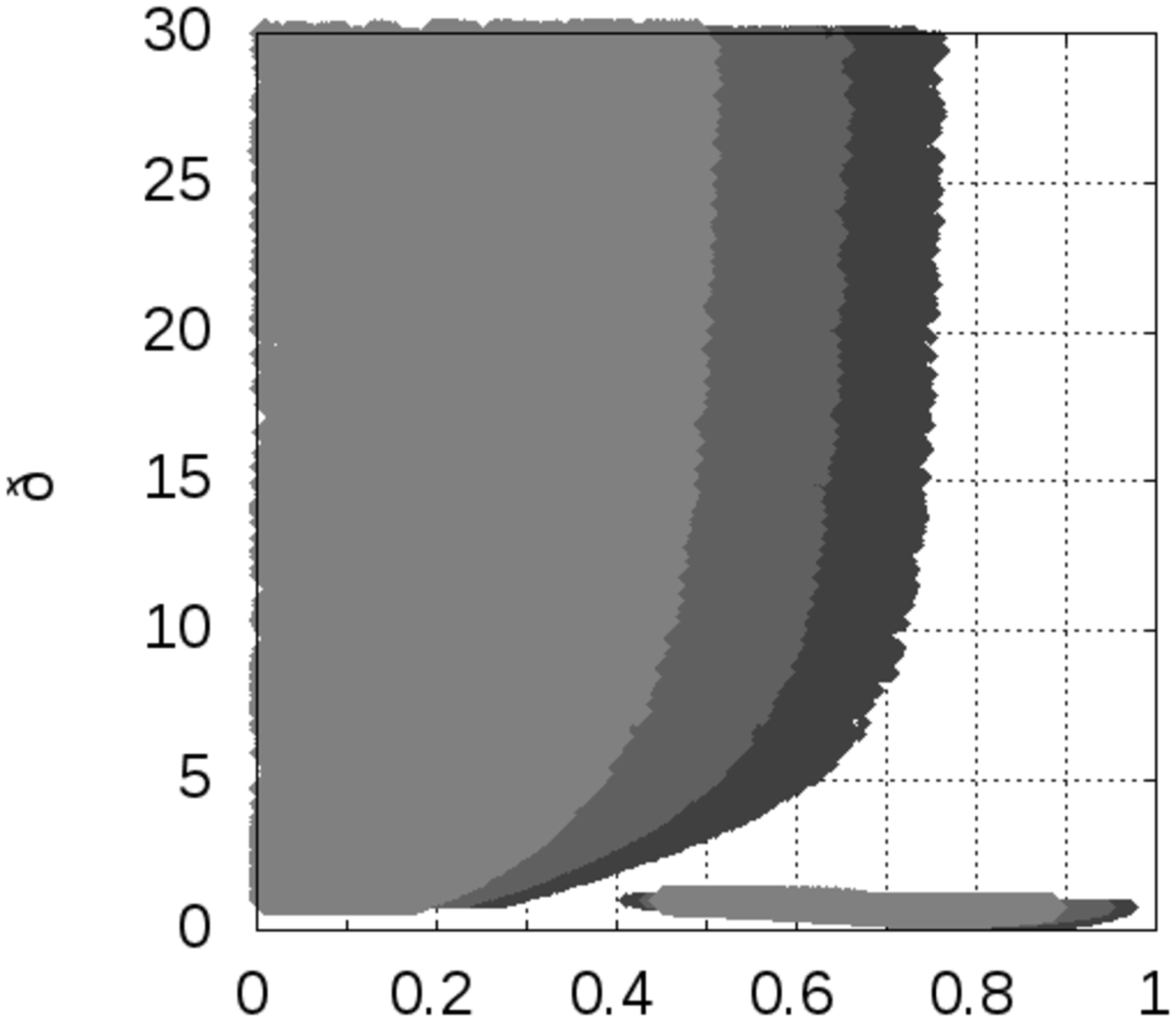}}} &
      \resizebox{50mm}{!}{\rotatebox{-90}{\includegraphics{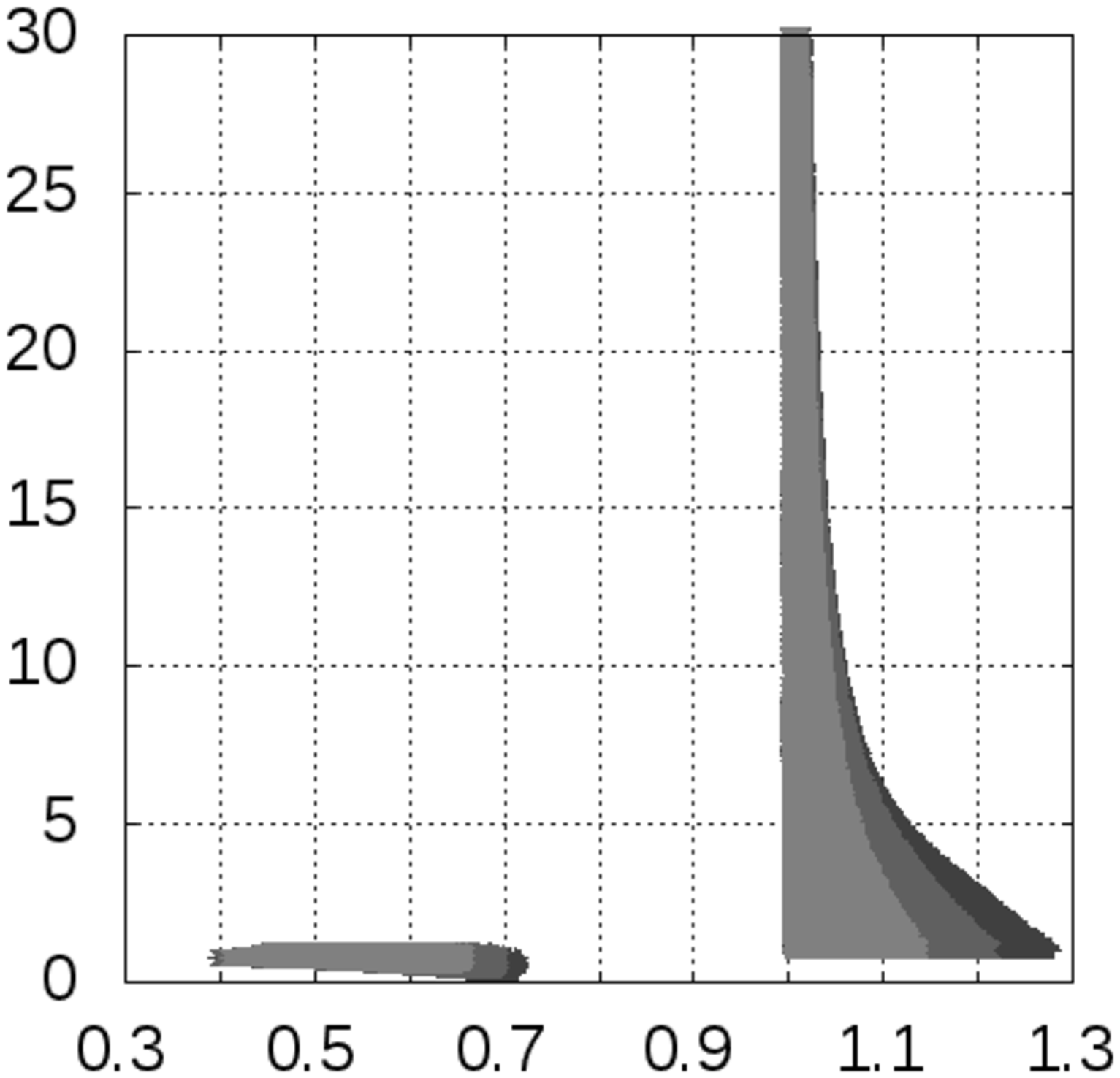}}} &
       \\

      \resizebox{50mm}{!}{\rotatebox{-90}{\includegraphics{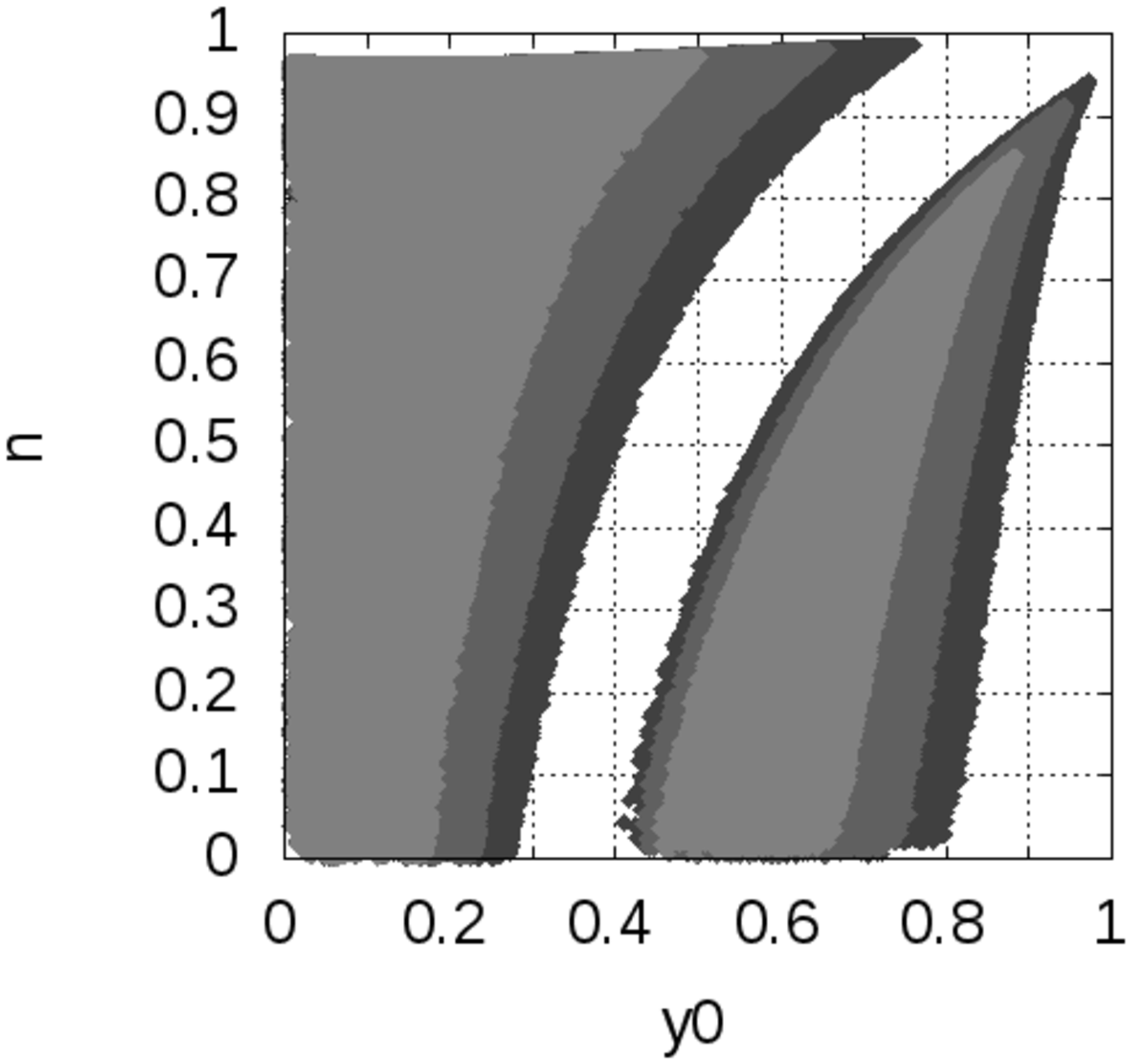}}} &
      \resizebox{50mm}{!}{\rotatebox{-90}{\includegraphics{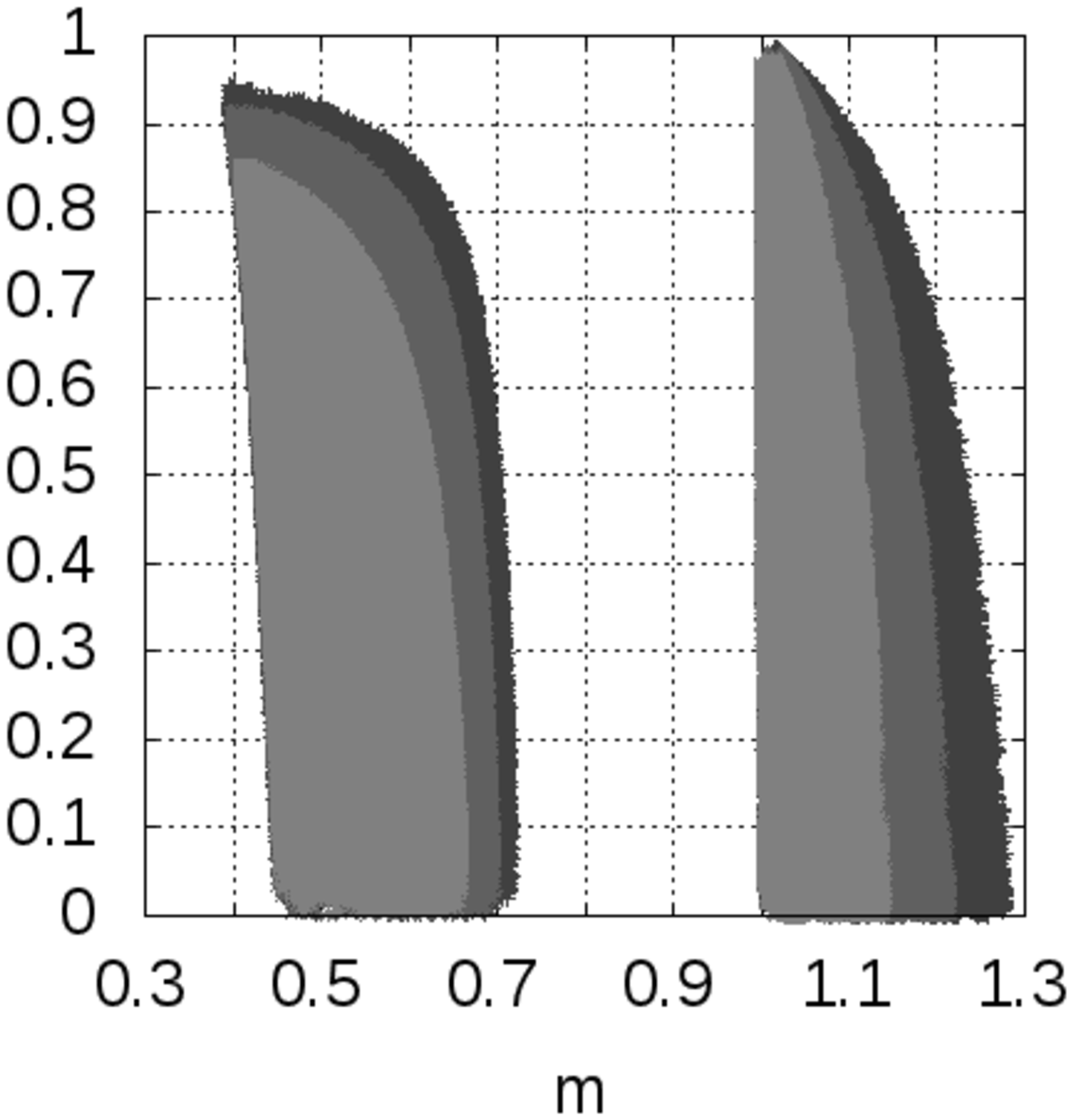}}} &
      \resizebox{50mm}{!}{\rotatebox{-90}{\includegraphics{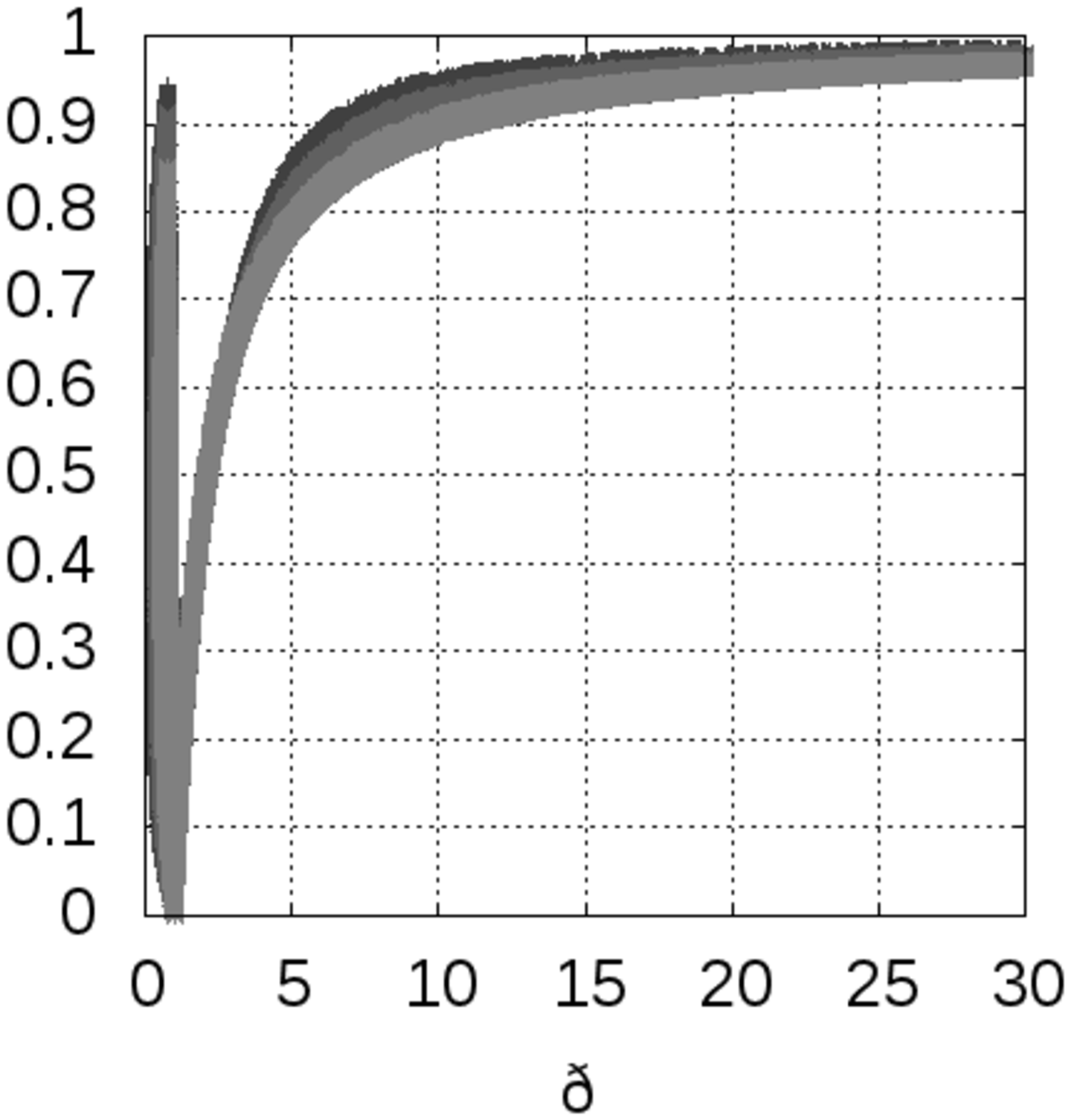}}} \\

    \end{tabular}

\caption{Marginalized contours for pairs of parameters are plotted. There are three confidence regions $68\%$, $95\%$, $99\%$ (from light grey to dark grey respectively) calculated for $\cal{A},\ \cal{R},$ and SN Ia jointly}\label{fig1}
\end{figure*}

\section{Results and Conclusions}\label{rac}
\setcounter{equation}{0}

Overall $\chi^2$ that has been used is
\be
\chi^2=\chi^2_{SN}+\chi^2_{R}+\chi^2_{A}
\ee

In Fig. \ref{fig1} we present contour plots showing the joint
marginal posterior distribution for each pair of FSF
model parameters. Each sub-panel shows three contours,
denoting roughly 68\%, 95\% and 99\% (from light gray to
dark grey respectively) credible regions.

In Fig. \ref{fig1} we see that there are two disjoined regions in the parameter space for which FSF singularities are allowed in the universe.
$\delta$ is always positive which is not surprising since only for positive $\delta$ there can be an accelerated expansion in the investigated model. Characteristic feature of the picture is that there are two qualitatively different regions divided by the value of the parameter $m$. There are two branches, the first for $m>1$, and the second for $m<1$. In the following tables we present ranges of the values of the parameters for three confidence levels. For the case when $m>1$:
\begin{center}
  \begin{tabular}{cccc }

 {\bf $1\sigma$ CL}     &    {\bf $2\sigma$ CL}    &  {\bf $3\sigma$ CL}        \\\hline
 $m\in(1.00,1.14)$      &  $m\in(1.00,1.21)$       & $m\in(1.00,1.28)$          \\\hline
 $\delta > 1.00$        &  $\delta > 1.00$         & $\delta > 1.00     $     \\\hline
 $n\in(0.0,0.97)$       &  $n\in(0.0,0.98)$        & $n\in(0.0,0.99)$           \\\hline
 $y_0\in(0.0,0.50)$    &  $y_0\in(0.00,0.66)$     & $y_0\in(0.00,0.76)$        \\

\end{tabular}
\end{center}

Below the case of $m<1$:

\begin{center}
\begin{tabular}{ccc}

 {\bf $1\sigma$ CL}     &    {\bf $2\sigma$ CL}    &  {\bf $3\sigma$ CL}        \\\hline
 $m\in(0.41,0.66)$     &  $m\in(0.39,0.67)$       & $m\in(0.39,0.72)$          \\\hline
 $\delta\in(0.52,0.99)$ &  $\delta\in(0.36,0.99)$  & $\delta\in(0.26,0.99)$     \\\hline
 $n\in(0.0,0.85)$       &  $n\in(0.0,0.91)$        & $n\in(0.0,0.94)$           \\\hline
 $y_0\in(0.45,0.88)$    &  $y_0\in(0.43,0.94)$     & $y_0\in(0.41,0.97)$        \\

\end{tabular}
\end{center}
\newpage
For the case of the branch $m>1$ the
maximum value at $3\sigma \ y_0$ parameter is $0.76$, which means that such a singularity can happen later than for the second branch $m<1$. While $y_0$ approaches $0.76$, $\delta$ grows stronger and as the value of $\delta$ more grows, $m$ becomes better constrained.

What seems to be important is that there is an allowed value of $m=2/3$ within $1\sigma$ CL, which could correspond to the dust filled Einstein-de-Sitter universe  in a close to big-bang limiting case. What is also interesting for this branch is that while the allowed values of the nonstandarcity parameter $\delta$ are small, i.e. $\delta<1$, the parameter $y_0$ approaches unity which means that the singularity may happen in the nearest future for this case. For $1\sigma$ CL, $y_0=0.885$ corresponds to $\sim 2\times 10^9$ years to the time of the singularity. For $3\sigma$ CL, $y_0=0.97$ corresponds to the present time at $\sim 0.37 \times 10^9$ years before the time of the singularity.\\
\indent In conclusion, we have shown that for a finite scale factor singularity there is an allowed value of $m = 2/3$ within $1\sigma$ CL,
 which corresponds to the dust-filled Einstein-de-Sitter universe for the close to big-bang limiting case.
The finite scale factor singularity may
happen within $2\times 10^9$ years in future for $1\sigma$ CL and its prediction at the present moment of cosmic
evolution cannot be distinguished, with current observational data,
from the prediction given by the standard quintessence scenario of
future evolution in the Concordance Model \cite{chevallier00, linder02, linder05, koivisto05, caldwell07, zhang07, amendola07, diporto07, hu07b, linder09}.
\section{Acknowledgements}

I warmly thank A. Balcerzak, M.P. D\c{a}browski, M.A. Hendry, and Yu.V. Shtanov for comments and discussions.\\
\indent I acknowledge the support of the {\it National Science Center grant No N N202 3269 40 (years 2011-2013).}}\\
\indent Part of the simulations reported in this work were performed using the HPC cluster HAL9000 of the Computing Centre of the Faculty of Mathematics and Physics at the University of Szczecin.

\end{document}